\documentclass[12pt,a4paper]{article}

\usepackage{subfigure}
\usepackage{graphicx}
\usepackage{epsfig}
\begin{document}

\baselineskip=20pt
\markboth
{\it S.K. Patra}
{\it S.K. Patra}
\mbox{ } \vspace{1.0in} \mbox{ }\\
\begin{center}
\begin{large}
{\bf New features in the stability and fission decay of superheavy Thorium
isotopes} 
\\[2ex]\end{large}
L. Satpathy, S.K. Patra and R.K. Choudhury\footnote{Present address:
Bhabha Atomic Research Centre, Mumbai, India}\\[2ex]

Institute of Physics, Bhubaneswar-751 005, India \\[2ex]

\end{center}

\bigskip
\begin{small}
\centerline {\bf Abstract:}

Superheavy isotopes are highly neutron rich nuclei in the vicinity of
neutron drip-line, stabilized by shell effect against the instability due
to repulsive component of nuclear force, analogous to superheavy
elements similarly stabilized against Coulomb instability.
Here we discuss the stability and fission decay properties of such
nuclei in the $^{254}$Th region and show that they are stable 
against $\alpha$ and
fission decay and have $\beta$-decay life time of several tens of seconds. In
particular, the $^{254}$Th nucleus has a low
fission barrier and unusally large barrier width. This makes it an ideal
thermally fissile nucleus, if formed by means of a thermal neutron,
like other known nuclei such as 
$^{233}$U, $^{235}$U, $^{239}$Pu in this 
actinide region. It shows a new mode of fast fission decay, which
may be termed as multifragmentation fission, in which in addition to 
two heavy fragments large number of scission neutrons are simultaneously produced.
Its likely synthesis during the r-process nucleosynthesis will have important
bearing on steller evolution, and here in the laboratory, it has great
potential in energy production.

\end{small}
\vfil
\eject

The superheavy elements in $Z= 114-126$ region \cite{she1,she2,she3} 
which are vulnerable to spontaneous
fission \cite{y1,y2,x1,x2,x3} due to the influence of repulsive Coulomb force of protons, 
have been predicted 
to be stable due to shell effect. Due to this effect, the periodic table
has extended upto Z=116 \cite{she1} by now with promise to extend further along
the valley of stabilitiy, thus elongating the stability peninsula\cite{ref1,ref2,ref3}.
Another effect that was shown by us earlier is the shell stabilisation of 
highly neutron rich superheavy isotopes of $Z=62, 78$ and 90 nuclei \cite{ref4}. 
It is known that the nucleon-nucleon force itself has a repulsive
component (triplet-triplet, singlet-singlet) whose contribution 
progressively increases with the increase of neutron 
number making the nucleus unstable. The 
shell effect can stabilize this instability giving rise to new magic
numbers in the vicinity of neutron drip-line. This is a complementary
parallel effect which should
widen the stability peninsula. In fact, in an extensive study involving
three different methods, infinite nuclear matter (INM) model 
\cite{ref5,ref5a,ref5b,ref5c}, relativistic
mean field (RMF) theory \cite{ref6,ref6a,ref7,ref7a,ref7b} and Strutinsky shell correction calculation
\cite{ref8}, it was
shown \cite{ref4} that 
islands of stability around new magic numbers N=164, Z=90; N=150, Z=78;
and N=100, Z=62 will exist giving rise to superheavy isotopes $^{254}$Th,
$^{228}$Pt and $^{162}$Sm. The N/Z ratios of these nuclei are 
1.82, 1.92 and 1.62
respectively, in contrast to the
value of 1.54 for the doubly closed shell superheavy element $^{298}X_{114}$ which
is quite
similar to that of $^{208}$Pb and $^{235}$U in the valley of stability. This result
is depicted here in Figure 1.
\begin{figure}[ht]
\epsfxsize=10cm
\begin{center}
\includegraphics[width=16cm,height=10cm,angle=0]{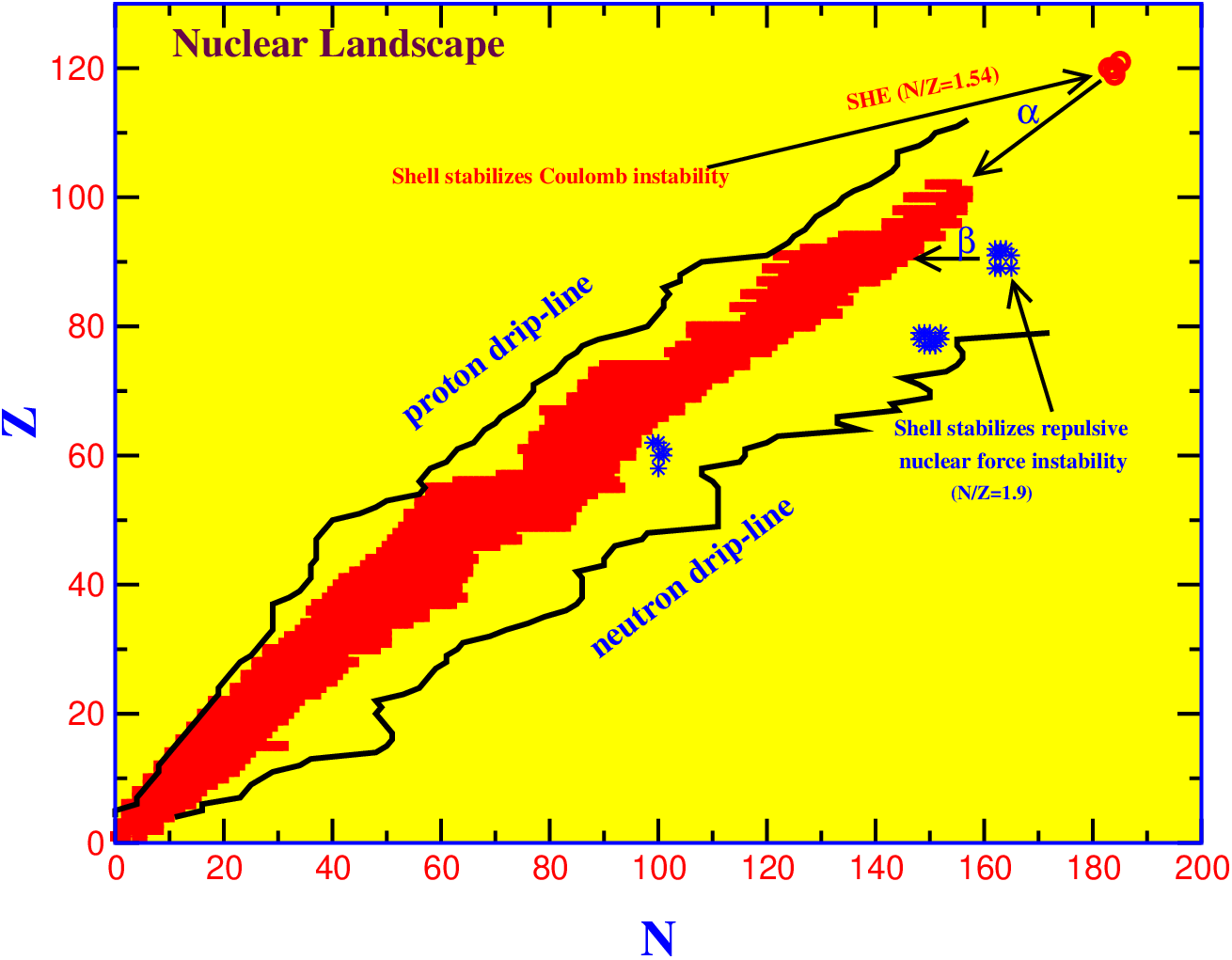}
\end{center}
\caption{The new islands of stability identified around the magic numbers
N=100, Z=62; N=150, Z=78 and N=164, Z=90
have been shown by the blue stars in the nuclear chart. The red area represents the nuclei
with experimentally known binding energies \cite{ref9}. The neutron and proton drip-lines
shown as solid curves are the predictions of infinite nuclear matter model.
The superheavy elements occur along the line of $\beta-$stability and
is shown by red circles,
whereas the superheavy isotopes occur close to the neutron drip-line.
The superheavy elements undergo $\alpha-$decay
and the superheavy isotopes $\beta-$decay to attain higher stability.
The shell effect stabilizes the superheavy elements against Coulomb
instability while it stabilizes the superheavy isotopes against
repulsive nuclear force instability.}
\label{fig1}
\end{figure}
It may be recalled that a new frontier in 1980's has opened up with
the production of radioactive ion beams with the prospect of
synthesis of about 5000 nuclei in the laboratory, and thereby, extending  the
nuclear peninsula upto drip-line region. The ultra neutron rich nuclei
are expected to show totally new properties and will probably be of immense
utility to mankind. Here we investigate the properties of a representative
ultra-neutron-rich nucleus $^{254}$Th which is a superheavy isotope of
thorium. It is known that $^{232}$Th is a fertile material whose fission 
decay properties are well known.
We show that $^{254}$Th is stable against $\alpha-$decay and
fission decay and has $\beta^-$decay
half-life of tens of seconds. Most interestingly, $^{254}$Th has unusual fission 
properties with low fission barrier of 3.57 MeV, but a very large
barrier-width, which makes it infinitily stable against spontaneous fission 
but fissionable if formed by means of a thermal neutron.
It will undergo a new mode of fission decay
in which, in addition to two fragments, large number of surplus neutrons will be
instantaneously produced. This may be termed as multi-fragmentation fission.
Due to highly stable character, this $^{254}$Th nucleus will have  important 
implications for $r-$process nuclear synthesis
in stellar evolution and also great potential for energy production
in the laboratory.

\begin{figure}[ht]
\epsfxsize=10cm
\begin{center}
\includegraphics[width=12cm,height=9cm,angle=0]{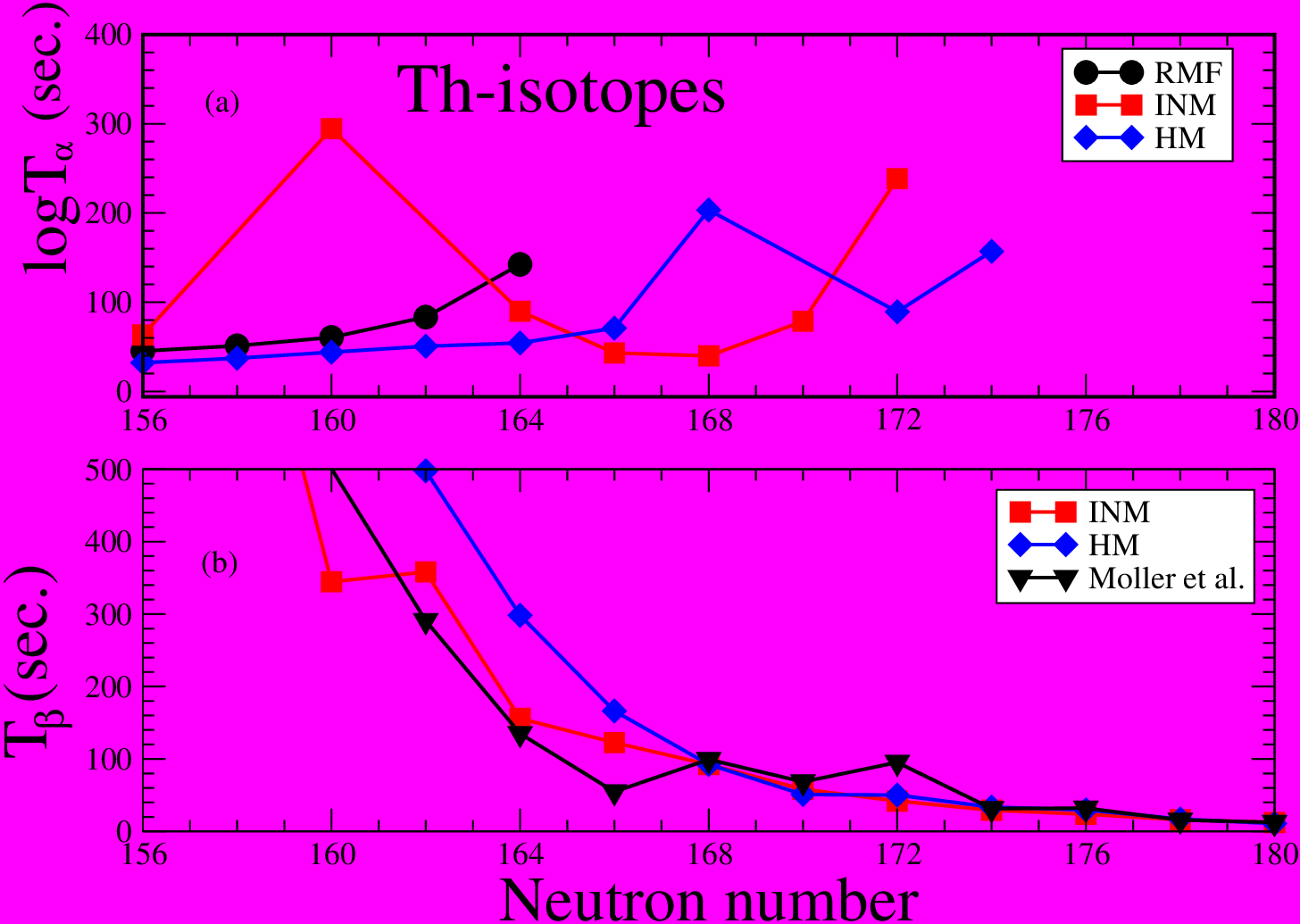}
\end{center}
\caption{(a) $\alpha-$decay half-life for Th isotopes using the empirical
formula $logT_{\alpha}=(aZ+b)Q_{\alpha}^{-1/2}+(cZ+d)$ of Viola and
Seaborg \cite{ref10}. The $Q_{\alpha}-$value calculated by
using the mass prediction of INM, RMF and HM models. The $\alpha-$decay
half-lives obtained by all the three models show similar
trend. All the three models show a peak around $N=160-170$ before 
$Q_{\alpha}$
becomes negative with increase of neutron number forbidding the
process. From the y-axis it is clear that the Th isotopes are
extremely stable against $alpha-$decay.
(b) The $\beta-$decay half life for Th isotopes
using the formula \cite{ref11,ref11a} 
$T_{\beta}=\frac{18\times 10^5\triangle ln 2}{W_0^6}$ sec. is calculated,
where $\triangle=$number of states of the daughter nucleus
within 1 MeV of ground state nucleus $\times$
$exp(-A/290)$ with A being its mass number and
$W_0=BE(Z+1,A)-BE(Z,A) + 1.29$ MeV.
The value of the $\beta-$decay half-lives decreases almost exponentially
with increase of neutron number. $^{254}$Th has an appreciable life-time of 
tens of seconds.
}
\label{fig2}
\end{figure}

\begin{figure}[ht]
\epsfxsize=10cm
\begin{center}
\includegraphics[width=12cm,height=10cm,angle=0]{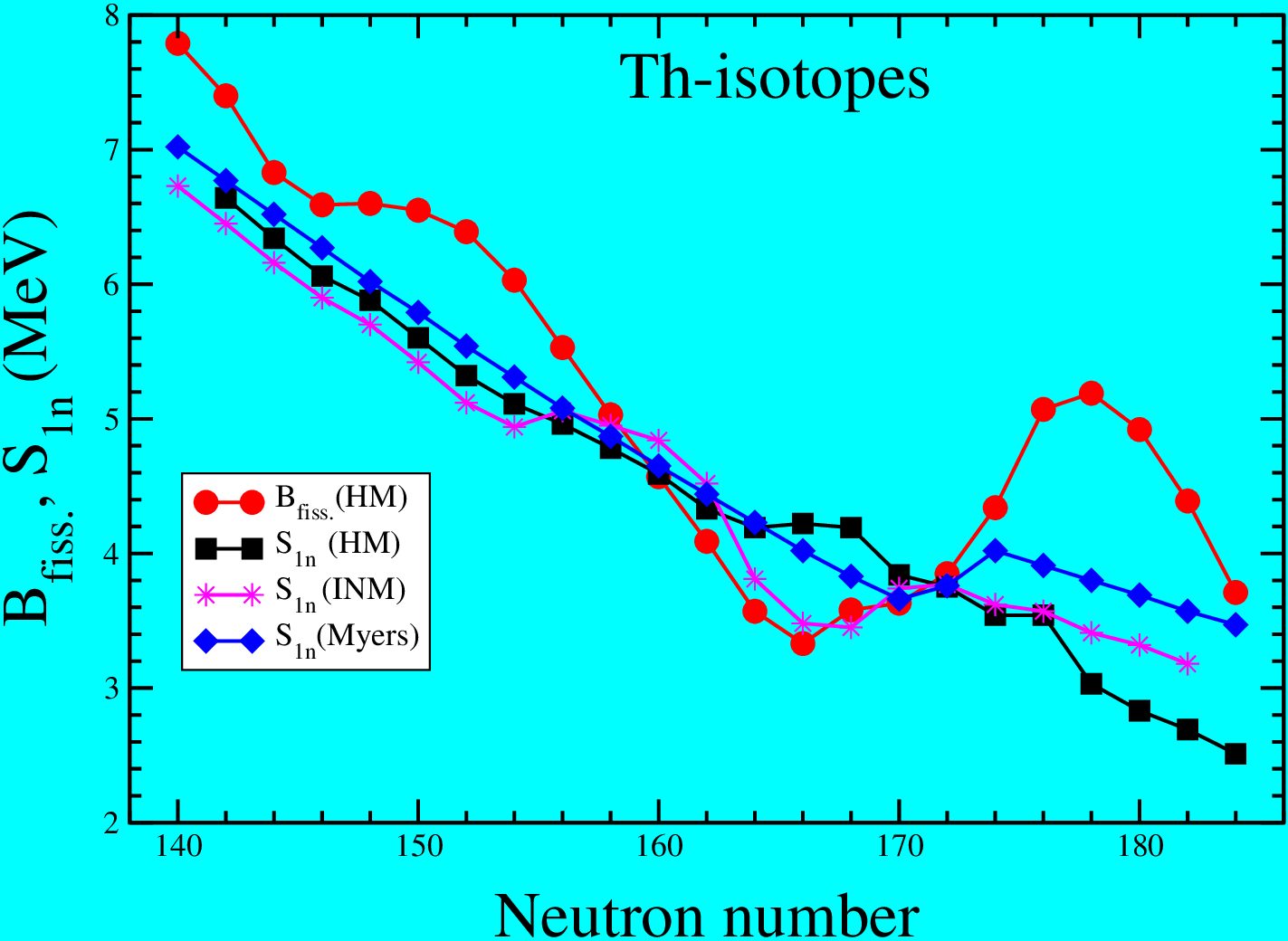}
\end{center}
\caption{Fission barrier $B_f$, and binding energy of the last
neutron $S_{1n}$ as a function of mass number A for Th isotopes.
The $B_f$ are taken from Howard and M\"oller \cite{ref12} and $S_{1n}$ 
are taken from Refs. \cite{ref12},
\cite{ref13} and \cite{ref14} for HM, INM and droplet model, respectively.
The neutron separation energy predicted by all the three models
predict a general pattern. In all the three cases the fission barrier lies
above the single particle energy, except for the region $N=162-170$, which
will be vulnerable to thermal neutron fission.}
\label{fig3}
\end{figure}

\bigskip
\bigskip

\noindent {\bf Alpha and Beta decay of Th isotopes}
\bigskip

The usual magic nuclei with nucleon numbers 2, 8, 20, 50, 82 and 126 
have their special properties, in particular extra stability, 
due to the close shell, 
and they occur along the central line
in the valley of stability.  However, as one goes across the valley
towards the neutron drip line, as is shown in Figure 1, new shell closures can arise giving
rise to new magic nuclei, and new islands of stability around them. The 
search and discovery of such nuclei is on the frontier of nuclear physics
research today. It will be interesting to study the decay properties
of the ultra neutron rich heavy nuclei. It is well known that, 
in general, heavy nuclei with $A>150$ are unstable against $\alpha-$decay
while for lighter nuclei this process is improbable. Infact the
superheavy nuclei $^{289}X_{114}$, $^{292}X_{114}$ show such decay, which
seems to be the case for all heavy nuclei along the line of stability.
The heavy neutron rich nuclei are expected to show deviation from this
feature.  In Figure 2a, the $\alpha-$decay half-lives of different Th
isotopes calculated
using the predictions of the masses in INM \cite{ref13} and 
Howard-M\"oller (HM)\cite{ref12} mass models and RMF
theory \cite{ref15} are shown. 
The calculation of $\alpha-$decay half-life $T_{\alpha}$ has been 
done using the widely used standard formula
$logT_{\alpha}/sec.=(aZ+b)Q_{\alpha}^{-1/2}+(cZ+d)$, where $a=1.66175$, $b=-8.5166$,
$c=-0.20228$ and $d=-33.9069$, due to Viola and
Seaborg \cite{ref10}. The value of $Q_{\alpha}$ is estimated by using the
standard relation $Q_{\alpha}(Z,N)=BE(Z-2,N-2)+BE(2,2)-BE(Z,N)$, 
where BE is the binding energy.
In general, all the isotopes are quite stable against $\alpha-$decay
with decay life-time exceeding 10$^{20}$years for all the three mass models.
The RMF and INM show a prominent peak around $N=160\sim 168$ as mark of
extra stability for isotopes around N=164. Beyond that
the $Q_{\alpha}$ values become negative prohibiting such decay altogether.
Thus $^{254}$Th can be considered to be stable against $\alpha-$decay.

The neutron-rich nuclei are generally $\beta^-$ active. They will 
successively emit $\beta^-$particles until they reach the line of $\beta-$stability
in the valley, a parallel scenerio to the superheavy elements which attend
stability by successive $\alpha-$emission (see Figure 1). The $\beta-$decay half-lives
$T_{\beta}$ are estimated by using the standard formula due to Seeger, Fowler and Clayton
\cite{ref11a} $T_{\beta}=\frac{18\times 10^5\triangle ln 2}{W_0^6}$ sec., 
where $\triangle=$number of states of the daughter nucleus 
within 1 MeV of ground state nucleus $\times$
$exp(-A/290)$ with A being its mass number and
$W_0=BE(Z+1,A)-BE(Z,A) + 1.29$ MeV. The half-lives of Th isotopes so 
obtained are presented in Figure 2b, which stop falling rapidly
from N=164 with some flat peaks thereafter indicative of extra
stability, and the values are  in the realm of tens of seconds. 
Thus $^{254}$Th can be considered to be a 
fairly stable nucleus for laboratory studies. 

\bigskip
\bigskip

\noindent {\bf Fission decay of $^{254}$Th}

\bigskip

The study of the fission decay properties of $^{254}$Th is of special
importance because the lighter stable isotope $^{232}$Th is known to be
a fertile nucleus. The question how the ultra-neutron-rich
superheavy isotopes for the actinide nuclei will decay by fission
has not been addressed before. It is here one may expect some exotic
new phenomena. The fission barrier determines the fission decay properties of
the nucleus. The nucleus can overcome the barrier and undergo decay 
when excited by an
external agency like an energetic neutron incident upon it. The neutron 
separation energy $S_n$ provides a measure of excitation for the fission decay process.
We have plotted in Figure 3 the neutron separation
energy $S_{n}$ and fission barrier $B_f$ of the even Th isotopes as a
function of neutron number. The black squares and the green stars
are the $S_{n}$ values of the HM and INM models. They are
connected by lines to guide the eye. The closed red circles represent
the values of fission barrier $B_f$ taken from Howard and M\"oller 
\cite{ref12}. It is well known that if $S_{n} < B_f$, then the nucleus
can not undergo thermal neutron fission. The fission threshold
$E_{nm}=B_f-S_{n}$ has to be overcome by impiging with an energetic
neutron, and thereby the nucleus will undergo fission. Hence,
if $S_n > B_f$, then thermal neutron (with practically zero energy)
can cause fission. The thermal fission becomes a very attractive 
possibility for energy generation. It is interesting to see in
Figure 3 that, although normal Th ($^{232}$Th) is not thermally
fissile, its isotopes around N=164 in the range N=162$-$170
are thermally fissile, which has important implication and consequences.
We have as yet only a few generally known 
nuclei which are fissionable with
thermal neutron, out of which only one i.e., $^{235}$U 
is naturally occuring with an abundance of 0.71$\%$ in natural
uranium, and the others like $^{233}$U and $^{239}$Pu can
be artificially produced. Of course they have long half-lives of
$1.6\times 10^{5}$ and $2.1\times 10^{4}$ years respectively, but
$^{254}$Th as will be shown aposteriori is permanently stable
against spontaneous fission. 

\bigskip
\bigskip

\noindent {\bf Fission barrier profile of $^{254}$Th}

\bigskip

We now consider the nature of the fission
decay mode of $^{254}$Th, which is primarily governed by the
profile of the fission barrier. The height and width of the fission barrier
which is supposed to be parabolic in nature have to be obtained.
We have followed an empirical method to get the width of the barrier
from the systematics of the known experimental fission half-lives,
and extrapolate them to extremely neutron rich region of interest to us.
The fission half-lives can be calculated in a simplistic manner as
$\tau_{1/2}=ln 2/np$, where n is the number of
barrier assault by the decaying fragment, related to the width
$\hbar\omega$ by $n\hbar=\hbar\omega/2\pi$. And $p$ is the
penetrability of the barrier given by $p=[1+exp(2\pi B_f/\hbar\omega]^{-1}$.
Taking the values of $B_f$ from reference \cite{ref12}, and using the
experimental $\tau_{1/2}$ \cite{ref16}, we get the values of $\hbar\omega$, the systematics
of which so obtained are shown in the plot of $\hbar\omega$ versus
$B_f$ in Figure 4 for various actinide nuclei. It is indeed very 
revealing as well as interesting
that the plot shows a linear behaviour with 
progressively increasing slope with the increase of
proton number of the elements from 90 to 96. For the next element Cf with Z=98,
the linear behavior gets fuzzy. However, the mean follows the trend
with a higher inclination. The trend is more conspicuously restored
for Fm with Z=100. This deviation correlates well with the fission mass yield
systematics, where considerable deviation from the standard
well-defined two peaks occur for Fm isotopes. It may be mentioned that
such systematics of $B_f$ versus $\hbar\omega$ has been analysed for the
first time here. The width $\hbar\omega$ for any isotope with calculated fission 
barrier may be
obtained by extrapolation of the linear graph. For $^{254}$Th
with a fission barrier of 3.57 MeV, we obtained the value of
$\hbar\omega$ close to zero. Now we can construct the
parabolic barrier of base width $\triangle r$ somewhat schematically, using the value of
$\hbar\omega$ following the relation $1/\hbar\omega=d^2V/dr^2$, where
$V$ is the potential energy. The barrier so obtained is presented
as the blue curve in Figure 5, which is an extremely flat and wide barrier.
For comparison, we have presented in the same figure the
fission barrier of $^{232}$Th as red curve 
constructed with $\hbar\omega=0.415$ MeV obtained from
the experimental half-life \cite{ref16}, and $B_f=7.40$ MeV from Howard and
M\"oller \cite{ref12}. The unusually large fission width derived above in
case of $^{254}$Th is quite understandable and in confirmity with expectation.
It is not necessarily true that the fission barrier falls with
the increase of neutron number in an isotopic chain, it may
even rise \cite{aleks}. So the wide barrier obtained
here is rather specific to $^{254}$Th, which may be related to
its shell closure structure.

Our experience from the
fission studies of uranium, has shown that the excess neutrons
mainly inhabit the neck region when the nucleus deforms on its
journey to fission, which is related to the width of the barrier.
Since in the case of $^{254}$Th, 22 excess neutrons participate
compared to that in $^{232}$Th, a larger neck must be formed
giving rise to a flat and wider barrier. This feature is also in tune
with the fact that in the limiting case of no Coulomb force, the
barrier will disappear reminiscent of infinite width and zero
height. This interesting feature of the barrier of $^{254}$Th with low height
but large width points out a new feature where, the nucleus is
infinitely stable against spontaneous fission. But with a slight deposition of
energy with a thermal neutron, it will undergo fission. Thus the 
fission half-life of spontaneous fission is nearly infinite.

\begin{figure}[ht]
\epsfxsize=10cm
\begin{center}
\includegraphics[width=12cm,height=10cm,angle=0]{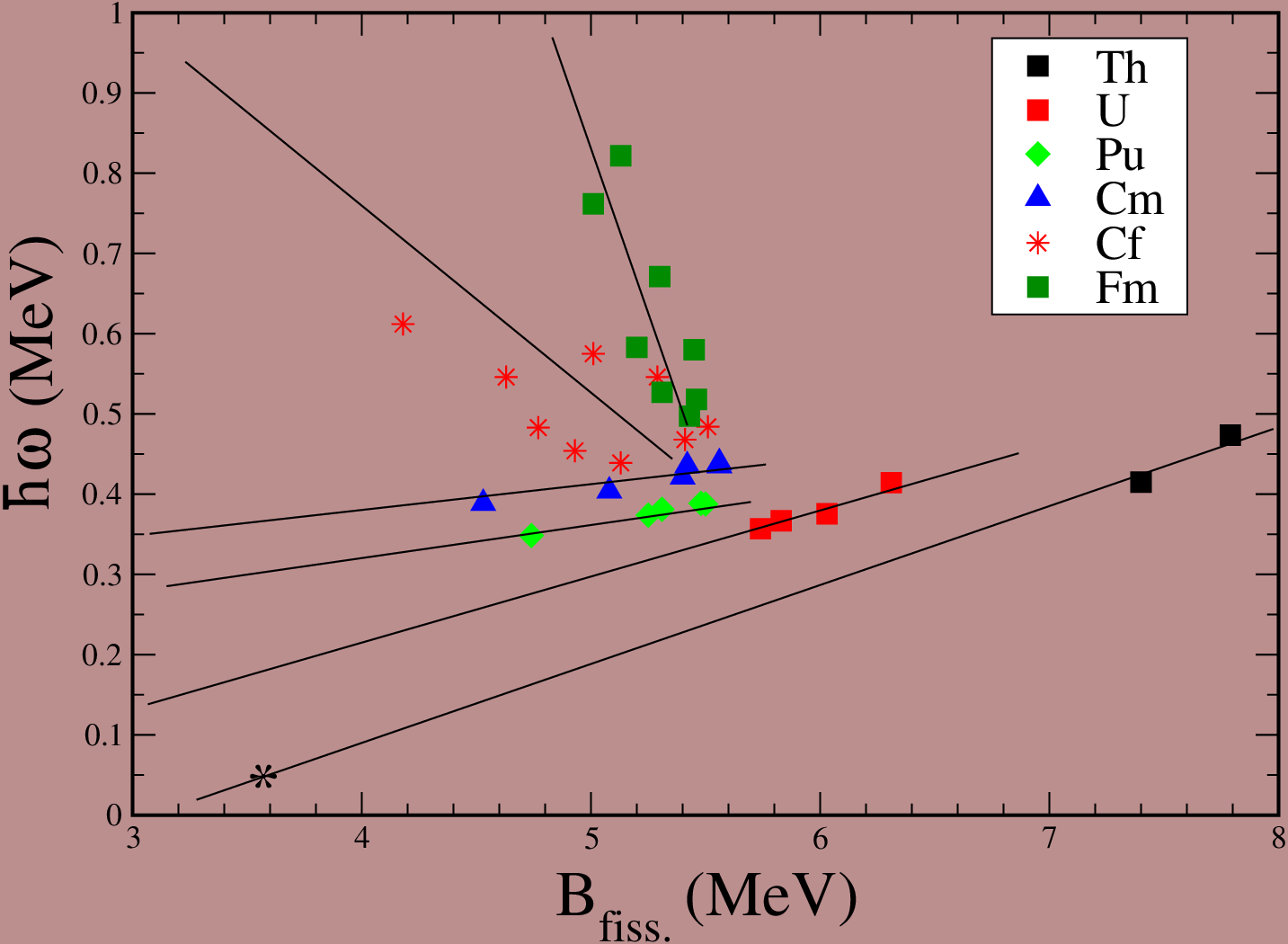}
\end{center}
\caption{The barrier width $\hbar\omega$ as  function of fission barrier $B_f$.
The $B_f$ values and the experimental half-lives taken from 
Howard and M\"oller \cite{ref12} and  Ref. \cite{ref16}, respectively
have been used to extract the barrier widths of all the known nuclei showing 
fission decay in the actinide region. This Figure shows that for each
element $\hbar\omega$ 
and $B_f$ show a linear relationship through a straight line behavior
depicted in the figure. The star on the line for Thorium in figure,
denotes the value of $\hbar\omega$ to be 0.05 MeV for its $B_f=3.57$ MeV.
This  linear relationship has been used to obtain $\hbar\omega$
for unknown isotopes by extrapolation shown by straight line.
}
\label{fig4}
\end{figure}

\begin{figure}[ht]
\epsfxsize=10cm
\begin{center}
\includegraphics[width=12cm,height=8cm,angle=0]{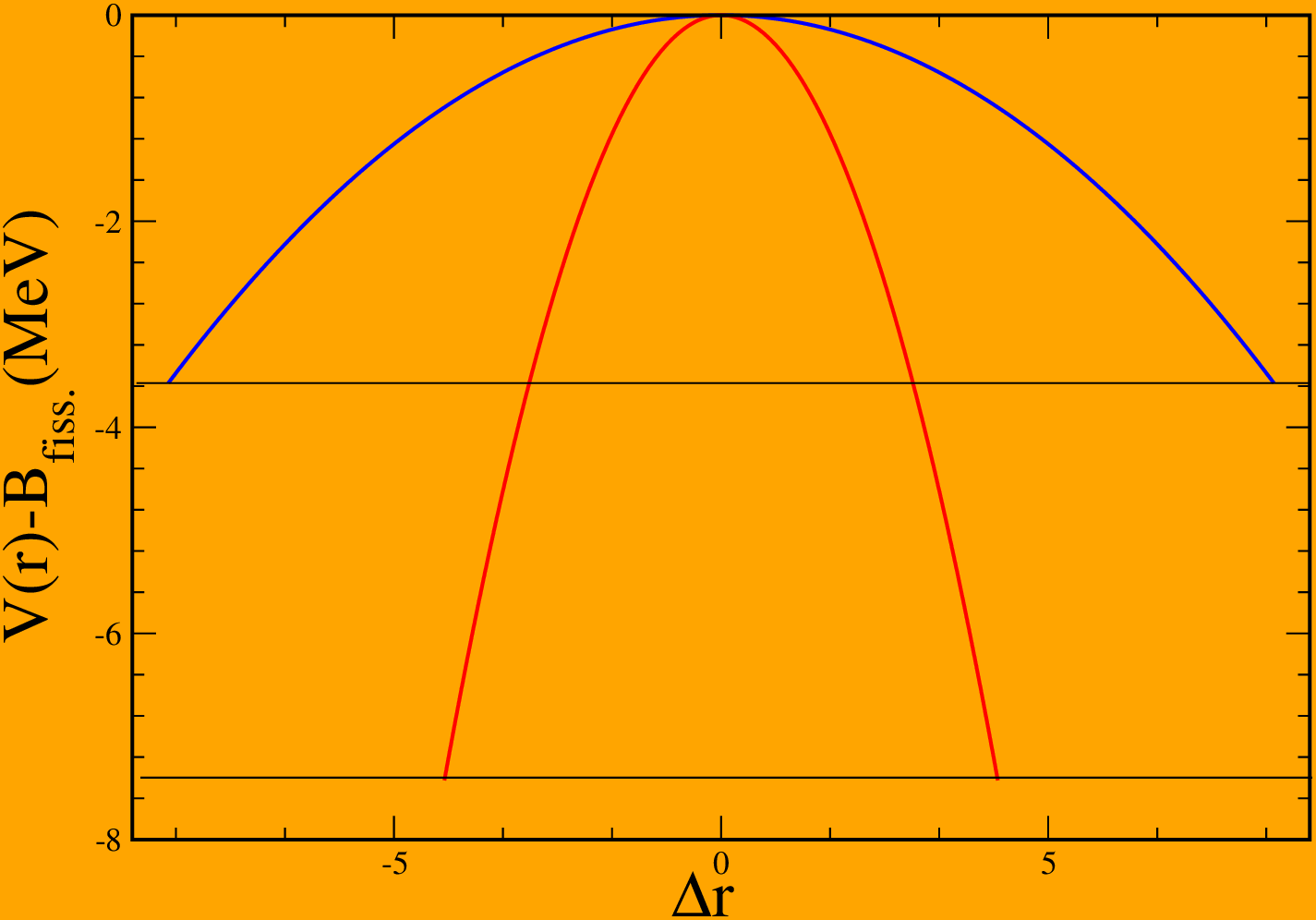}
\end{center}
\caption{The schematic curve is constructed using fission barrier
height $B_f$ and width $\hbar\omega$ to represent the 
shape of the parabolic fission barrier using the expression $y=-ax^2$. 
Here $a=\hbar\omega$, $y=d^2V/dr^2$ and $x=\triangle r$ the base
width of the parabola, where
$V$ is the potential energy. This figure shows that smaller the
value of $\hbar\omega$ larger is the width. The normal nuclei like $^{230,232}$Th
or $^{235,238}$U have $\hbar\omega\approx 0.4-0.5$MeV. In the figure, the
red curve represents the fission barrier of $^{232}$Th and the
blue one of $^{254}$Th. For the neutron rich
superheavy isotope of $^{254}$Th, the width is flattened unexpectably
and the height decreases considerably as whown in the blue curve. This
flattening of fission barrier makes the nucleus stable against
spontaneous fission decay, because of decreasing penetrability.
 On the other hand, due to the
decrease of fission barrier, with a minute inducement by a thermal neutron,
fission decay will occur.}
\label{fig5}
\end{figure}

\begin{figure}[ht]
\epsfxsize=10cm
\begin{center}
\includegraphics[width=15cm,height=10cm,angle=0]{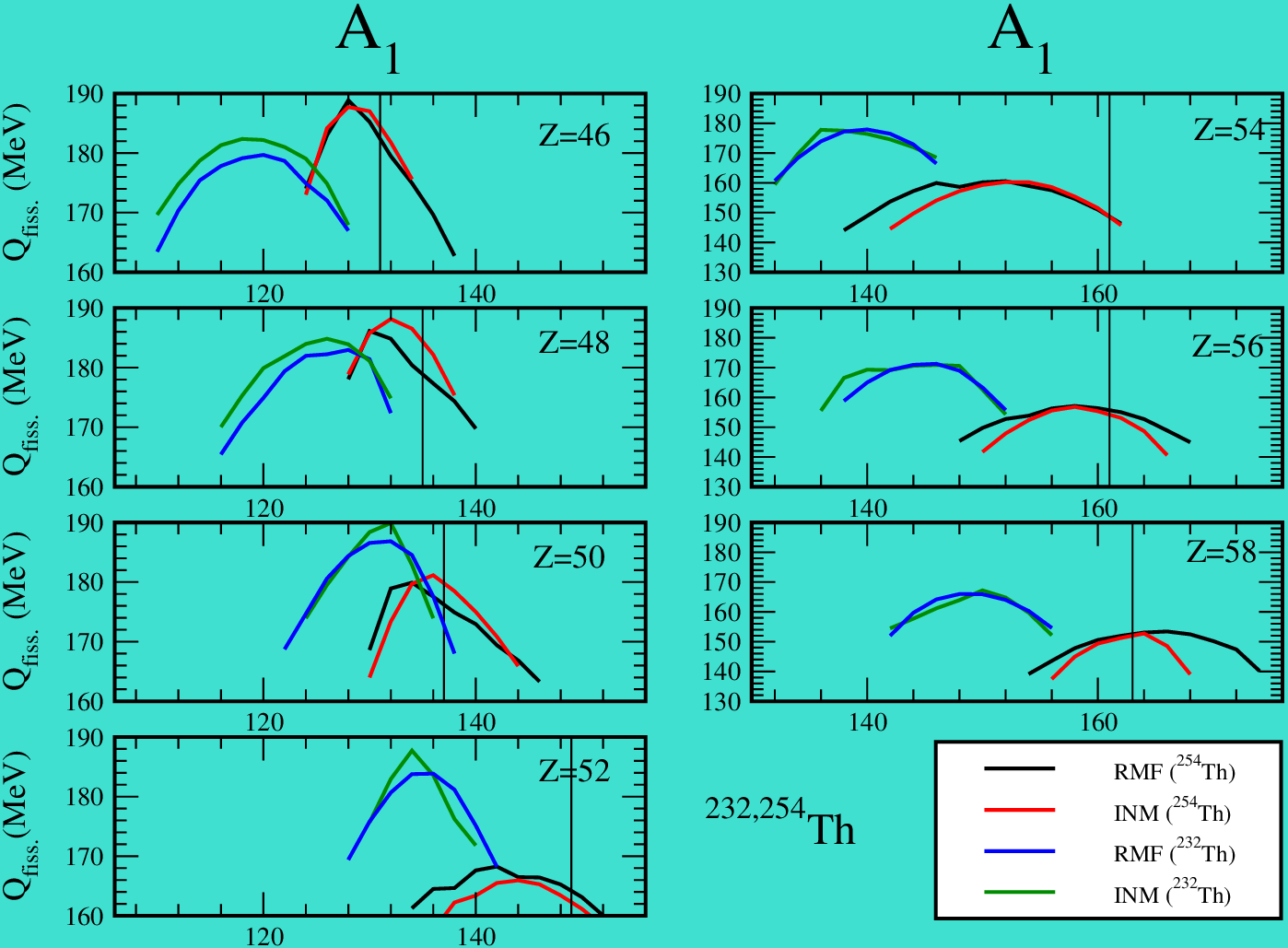}
\end{center}
\caption{$Q-$value distribution given by $Q^f=BE(A_1,Z_1)+BE(A_2,Z_2)
-BE(A,Z)$ for $^{232}$Th and $^{254}$Th as a function of $A_1$
fragment in the binary decay $A\rightarrow A_1+A_2$.
The binding energy used for calculation of $Q-$value is
taken from Ref. \cite{ref13} and \cite{ref15} for INM and
RMF respectively. The fission yield decreases drastically with increase or 
decrease of mass number of given element $(A_1,Z_1)$. Therefore, we have shown 
the distribution in the range 90 to 100$\%$ of the peak value in each case. 
The vertical line marks the
neutron drip-line for the corresponding element in each pannel.
From the figure, it is clear that for $^{232}$Th no fragment is 
produced with isotopes beyond the drip-line.
On the other hand, for $^{254}$Th a larger number of
fragments are predicted to lie away from the drip-line, giving rise
to some neutron emission simultaneously, which
may be termed as multi-fragmentation fission. 
}
\label{fig6}
\end{figure}

\bigskip
\bigskip

\noindent{\bf New mode of fission decay}
\bigskip

We next examine the decay mode of fission of a nucleus, ($A,Z$) decaying
to ($A_1,Z_1$) and ($A_2,Z_2$), solely guided by the
$Q-$value systematics of the process, where $Q$ is defined
as $Q^f(A,Z)=BE(A_1,Z_1)+BE(A_2,Z_2)-BE(A,Z)$. As is well known, 
the major driving force
for the decay  is the $Q-$value of the reaction. The probability of
fragment mass yield in a given channel is directly related to the
$Q-$value. In Figure 6, we have plotted the $Q-$values of the
binary decay into two fragments $A_1$ and $A_2$, as a function of the
mass number of the $A_1$ fragment, 
for all the relevant elements with even values of $Z$, starting
from 46 to 58.  The complimentary fragment ($Z_2$, $A_2$) is thereby fixed.
Since the yield falls rapidly 
with the decrease in $Q-$value for an element, we have only shown the 
distribution of $Q-$values lying 
above $90\%$ of the highest values. For the sake of comparison, the
$Q-$value distributions for both $^{232}$Th and $^{254}$Th are 
shown in Figure 6. The black and red curves pertain to $^{254}$Th
and blue and green to $^{232}$Th calculated using the predicted
masses in INM and RMF models respectively. 
Since $^{254}$Th is an ultra-neutron-rich
isotope, during its scission many neutrons will be left free as they
can not be bound to either of the fragments which may be already
on the drip-line. In the figure, vertical lines mark the
drip-line of the respective elements. As expected, the drip-line
falls nearly in the middle of the distribution
in case of $^{254}$Th, indicating that all the isotopes lying to the right of it will
be unstable against instantaneous neutron emission from the
fragments at scission. These drip-line neutrons will be
simultaneously produced along with the newly formed fragments.
In the usual fission process, neutrons are produced by the
emission from the fragments after they are fully accelerated. But in the
present case of $^{254}$Th, a certain number of neutrons will be simultaneously
produced, signalling a new mode of fission decay which may be
termed as {\it multi-fragmentation fission}. An order of magnitude of these 
prompt multi-fragmentation neutrons can be estimated from the mass yield plot for 
$^{254}$Th as shown in Figure 6, which turns out to be about
2 to 3 neutrons per fission. These are the additional neutrons apart
from the normal multiplicity of neutrons emitted from
the fragments. In case of neutron-induced fission of normal
$^{232}$Th and $^{235}$U nuclei, the fission neutron multiplicities are
of the order $2.3$ to $2.5$ \cite{ref17,ref18}. 
This number is, therefore, nearly doubled in case of
$^{254}$Th fission, which will have important implications on the
energetics of the fission process.

Although, the beta decay life-time of $^{254}$Th is few tens of seconds
it is much larger than the nucleon decay life time, which is 
of the order of $10^{-17}$ seconds and therefore it will have
a great implication in the $r-$process nuclear synthesis
and consequently stellar evolution. We may also conjecture that 
in the laboratory, such
nuclei can be of immense potential as energy source.

\bigskip
\bigskip
\noindent {\bf Conclusions}
\bigskip

In conclusion, we have studied the decay properties of Th isotopes, in particular
$^{254}$Th, a superheavy isotope of thorium which has been predicted earlier
to be a doubly close shell nuclei with N=164, and Z=90 stabilised by shell
effect against repulsive nuclear force instability. It is stable against 
$\alpha-$decay and has $\beta-$half-life of several tens of seconds. 
The fission decay properties of such neutron-rich nuclei in particular
in the actinide region have not been
addressed before, which has important bearing on the RIB programmes in many
laboratories in the world seeking the synthesis of about 5000 nuclei in the 
exploration upto the drip-lines. On the basis of the systematics of 
neutron separation energies
and fission barriers, it has been shown to have a rare property of being 
thermally fissile putting it into the rank of other such generally known 
nuclei like 
$^{233}$U, $^{235}$U and $^{239}$Pu. Its fission barrier profile
determined empirically using the systematics of experimentally
known so far in this actinide region 
fission half-lives and barrier of the actinide nuclei, shows it
to be very flat and wide. Such feature is in confirmity with the    
expected long neck due to the excess number of neutrons to the tune of 22 over that of the
$^{232}$Th, presumably supported by its shell closure property.
Coupled to this, its small barrier of 3.57 MeV, makes it
extremely stable against spontaneous fission, but highly 
vulnerable to thermal neutron fission, a quite unique property indeed. 
A new mode of fission decay named multifragmentation fission is 
predicted for this nucleus, where in addition to two heavy fragments, 
2 to 3 scission neutrons will be simultaneously produced. This
is in addition to the normal multiplicity of neutron emitted by the 
fragments which are of the order of 2.3 to 2.5. Thus the  doubling of the
neutron emission per fission will have strong implications as for the $r-$process 
nucleosynthesis in the steller evolution. It also presents an attractive
possibility as a source of energy production in the laboratory.

\vfil
\eject

\hfil\break
\vfil\eject
\end{document}